\documentstyle[aps,prl]{revtex}
\input epsf
\begin{document}
%\narrowtext
\title 
 {Coherent transport and nonlocality in mesoscopic SNS junctions:   anomalous magnetic interference patterns}
\author{Victor Barzykin$^{1,2}$ 
 and Alexandre M. Zagoskin$^2$
 \thanks{Email: zagoskin@physics.ubc.ca}}
\address{$^1$
National High Magnetic Field Laboratory,
1800 E. Paul Dirac Dr., Tallahassee, Florida 32310, USA;\\
$^2$Physics and Astronomy Department, The University of British Columbia, 6224 Agricultural Rd., Vancouver, B.C., Canada V6T 1Z1}\maketitle

\begin{abstract}
We show that   in ballistic  mesoscopic SNS junctions the period of critical current vs. magnetic flux dependence  (magnetic interference pattern),
$I_c(\Phi)$, changes
{\em continuously and non-monotonically}  from $\Phi_0$ to 
$2\Phi_0$ as the 
length-to-width ratio of
the junction grows, or temperature drops. In diffusive mesoscopic junctions the change is even more drastic, with the first zero of $I_c(\Phi)$  appearing at $3\Phi_0$.

The effect is a manifestation of nonlocal relation between the 
supercurrent density and
superfluid velocity in the normal part of the system, with the characteristic
scale $\xi_T = \hbar v_F/2\pi k_BT$ (ballistic limit) or
$\tilde{\xi}_T = \sqrt{\hbar D/2\pi k_BT}$ (diffusive limit), the normal metal coherence length,
and arises due to  restriction of the  quasiparticle phase space 
near the lateral boundaries  of the junction. It explains
the  $2\Phi_0$-periodicity  recently observed
by Heida et al. (Phys. Rev. B {\bf 57}, R5618 (1998)).
We obtained explicit analytical expressions for the
magnetic interference pattern   for a junction with an
arbitrary length-to-width
 ratio. 
Experiments are proposed to directly
observe the $\Phi_0\to 2\Phi_0$- and $\Phi_0\to 3\Phi_0$-transitions.
\end{abstract}

It is well established that the electrodynamics of the superconductors is
 nonlocal on the scale of
$\xi_0$ \cite{Pippard,AGD}, and so is the current-phase relation
(since vector potential and superconducting phase 
both enter through one gauge-invariant combination, superfluid velocity).
 In the normal layer of an SNS system
the place of $\xi_0$ is taken by the normal metal coherence length,
$\xi_{T} = \hbar v_F/2\pi k_B T$ (in the ballistic limit,
$L\ll l_i$, where $l_i$ is the impurity scattering length),
or   $\tilde{\xi}_T = \sqrt{ D/2\pi k_BT}$, (in the diffusive limit, $l_i \ll L$)
where $D = v_F l_i/3$ is the diffusion constant of electrons in the dirty metal \cite{Kulik,Swidzinskii}.    This length can obviously
  greatly exceed $\xi_0$ at low enough temperatures, and is limited only by the inelastic scattering length in the system, $l_{\phi}$, which also diverges as $T\to 0$. (The latter  gives the scale of {\em classical} nonlocality in mesoscopic systems, related to formation of essentially nonequilibrium states, when currents are determined by potential drops far from the observation point, like in classical point contacts or Landauer wires.) In wide SNS junction, the nonlocality is responsible for the formation of current-carrying  Andreev
levels, formed in the normal part of the system by Andreev reflections
of quasiparticles 
from the off-diagonal scattering potential (order parameter) of the
superconducting banks
\cite{Kulik,Swidzinskii,Bardeen}, which provide the mechanism for the Josephson current flow.
In  mesoscopic SNS systems it is revealed in such effects as e.g.    conductance oscillations in Andreev interferometers(Ref.\cite{BLOM}
and references therein). It came nevertheless  
as a surprise, when   Heida et al. \cite{Heida} reported that the critical current vs. 
magnetic flux dependence (magnetic interference pattern) in  mesoscopic   junctions, formed by Nb electrodes in contact with 2D electron
gas of InAs heterostructure (S-2DEG-S junction),
 has a period $2 \Phi_0 = hc/e$, instead of the
standard $\Phi_0$ \cite{Barone}.  The authors of
\cite{Heida}   posed a question whether this unexpected result is due to 
a nonlocal superconducting current-phase relation specific for their narrow
(length-to-width ratio $ L/W \sim 1$, instead of usual $L/W \ll 1$), ballistic,
mesoscopic   junction. 
They   also asked whether
the diffusive character  of scattering at NS boundaries (which was
 prominent in the experimental samples) is important for  establishing
such nonlocality. A heuristic formula proposed in \cite{Heida}
successfully reproduced the periodicity, but not the shape of the 
observed magnetic interference pattern.

                                            Since the nonlocality under question is already present in wide SNS junctions,   where the
 magnetic interference
pattern is nevertheless $\Phi_0$-periodic, the problem clearly requires a more thorough consideration.  For example, the dephasing and inelastic scattering processes cannot be responsible
for the "wide-narrow`` difference, because the standard theory \cite{Swidzinskii,SAB} predicts  $\Phi_0$-periodicity 
 even when  $l_{\phi},\xi_T = \infty$.

In this paper we show
 that $\Phi_0\to 2\Phi_0$-transition  is  a size effect,
 resulting from the restrictions on allowed 
quasiclassical trajectories of quasiparticles in the normal layer near its
lateral edges. It is present already in the absence of diffusive Andreev
scattering (perfect NS boundaries). As the $L/W$ ratio drops, relative
contributions from the edges diminish, and in the limit $L/W\to 0$ 
the standard $\Phi_0$-periodicity is restored. The latter is actually
a result of certain contributions to the current cancelling each other,  
 not of their decay because of junction size exceeding the inelastic
scattering length.  We also predict  even more drastic effect  in the diffusive SNS junction, that can be qualitatively explained along the same lines.

\begin{figure}
\epsfxsize=5 in
\epsfbox{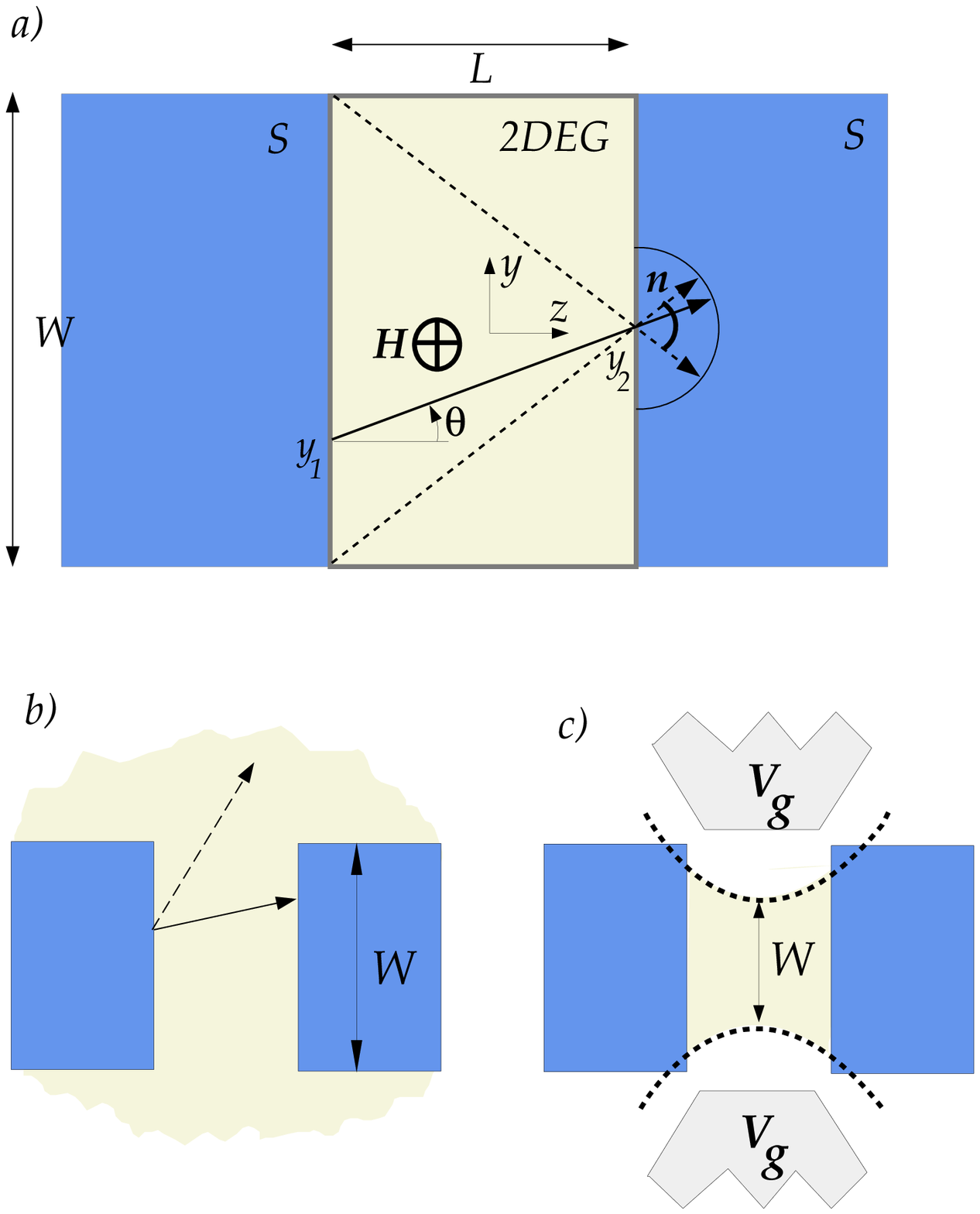}
\caption{(a) Josephson current in a ballistic S-2DEG-S junction.
(b) "Open'' SNS junction. Only the quasiparticle trajectories linking both superconductors
contribute
to the Josephson current. (c) Suggested experiment  for direct observation of
the $\Phi_0\to 2\Phi_0$-transition. The width  of
the system is changed continuously by the gate voltage, $V_g$.}\label{figY}
\end{figure}

 An important consequence of the "geometric``  origin of the effect is the 
continuous transition of the periodicity between $\Phi_0$ and $2\Phi_0$ as the
$L/W$ ratio changes. Indeed, the restriction on
 contributing   quasiparticle trajectories can be considered as having a
smaller effective flux through the system. The effective flux, and therefore the observed periodicity, obviously will be a {\em continuous}
 function of $L/W$. This transition could be directly observed in a   modification of the experimental setup of \cite{Heida}, where
the width of the channel is determined by the voltage applied to gate electrodes (Fig.\ref{figY}c). Such split-gate technique 
 is now widely used in hybrid  systems 2DEG-superconductor
(see e.g. \cite{Takayanagi}). Another experimental 
possibility, applicable to the junctions of fixed dimensions (like in
\cite{Heida}), will be discussed later.

First consider a ballistic S-2DEG-S junction (Fig.\ref{figY}a) in the limit 
$L \gg \xi_0$, assuming perfect Andreev reflections at NS interfaces, and
  totally absorbing lateral boundaries (at $y=\pm W/2$). The latter
condition, strictly speaking,
  corresponds to an "open'' SNS junction (Fig.\ref{figY}b), 
analogous to the structures investigated experimentally in \cite{Dimoulas}.
In case of laterally restricted 2DEG one should use, e.g., condition of
mirror  reflection from the side walls. 
We will see that this does not
  qualitatively change the results. 

The magnetic field   and   the vector potential
(in the Landau gauge) are
\begin{eqnarray}
{\bf H} = H \hat{\bf x},\:\:\:
{\bf A} = H y \hat{\bf z}.\label{gauge}
\end{eqnarray}
 Finally, we use the standard
steplike approximation for the superconducting order parameter in the banks
of the junction: 
\begin{eqnarray}
\Delta({\bf r}) = |\Delta_0|e^{i\phi_1}\theta(-L/2-z) + 
  |\Delta_0|e^{i\phi_2}\theta(z-L/2).
\end{eqnarray}
(In the gauge (\ref{gauge}) the superconducting phase is not affected by the 
field.)

  The Fermi wavelength in the normal part of the system,
$\lambda_F = h/p_F\: (\approx 0.1\:\mu$m), 
is small compared to all other length scales in the problem. 
This allows us to apply the
  method of quasiclassical Green's functions
integrated over energies \cite{Eilenberger}, 
\begin{eqnarray*}
g_{\omega}({\bf n},{\bf r}) = \frac{i}{\pi}
\int_{-\infty}^{\infty} d\!\xi{\cal G}_{\omega}(\xi,{\bf n}; {\bf r});\:\:\:
f_{\omega}({\bf n},{\bf r}) = -\frac{1}{\pi}
\int_{-\infty}^{\infty} d\!\xi{\cal F}_{\omega}(\xi,{\bf n}; {\bf r});\\
f_{\omega}^+({\bf n},{\bf r}) = f_{\omega}(-{\bf n},{\bf r})^*;\\
g_{\omega}({\bf n},{\bf r})^2 + f_{\omega}({\bf n},{\bf r})
f_{\omega}({\bf n},{\bf r})^+ = 1.
\end{eqnarray*}
Here ${\cal G}_{\omega}(\xi,{\bf n}; {\bf r}), 
{\cal F}_{\omega}(\xi,{\bf n}; {\bf r})$ are spatial Fourier transforms of
normal and anomalous Gor'kov functions of arguments ${\bf x}_{1,2} = 
{\bf r}\pm\delta{\bf r}$.  The momentum canonically conjugated 
to  $\delta{\bf r}$ is $${\bf p} = \sqrt{2m^*\xi}{\bf n}.$$  

The supercurrent density is then found as a sum over Matsubara
frequencies,
\begin{eqnarray}
{\bf j}({\bf r}) = - 4\pi ie{\cal N}(0)v_F T \sum_{\omega>0}\left<{\bf n}
g_{\omega}({\bf n},{\bf r})\right> =  
- 4\pi ie{\cal N}(0)v_F T \sum_{\omega>0}\left<{\bf n}
(g_{\omega}({\bf n},{\bf r}) - g_{\omega}(-{\bf n},{\bf r}))\right>_{n_z>0};
\end{eqnarray}
the angular brackets denote angular averaging over the Fermi surface.

Functions $g_{\omega}, f_{\omega}$ satisfy the set of Eilenberger equations,
that in the ballistic limit and in the absence of
the field is\cite{KO}
\begin{eqnarray}
(2\omega + v_F{\bf n}\cdot\nabla)f_{\omega}({\bf n},{\bf r}) = 
2 \Delta({\bf r}) g_{\omega}({\bf n},{\bf r});\\
v_F{\bf n}\cdot\nabla g_{\omega}({\bf n},{\bf r}) = 
\Delta({\bf r})^* f_{\omega}({\bf n},{\bf r}) - 
\Delta({\bf r}) f_{\omega}^+({\bf n},{\bf r}).
\end{eqnarray}
Its characteristics are   straight lines,   interpreted as
quasiclassical trajectories of quasiparticles\cite{KO}.
 It should be solved  in three regions: $z<-L/2, |z|<L/2, z>L/2$, using the limiting 
values in the bulk of left, right superconductor,
\begin{eqnarray}
f_{\omega}({\bf n},z=\mp\infty) = \frac{|\Delta_0|e^{i\phi_{1,2}}}
{\sqrt{|\Delta_0|^2
+ \omega^2}};\\
g_{\omega}({\bf n},z=\mp\infty) = \frac{\omega}{\sqrt{|\Delta_0|^2
+ \omega^2}},
\end{eqnarray}
and then matching solutions over the interfaces. Eventually, this will yield
the standard sawtooth expression for the superconducting current density
in an SNS junction\cite{SAB,Bardeen,Ishii}, which we will write   
in the form \cite{Zagoskin} for a point at the right NS boundary:
\begin{eqnarray}
j_z({\bf r}= y_2 \hat{\bf y}+(L/2) \hat{\bf z}) = \int_{\cos\theta>0}  d\!\theta\frac{ev_F\cos\theta}{\lambda_F W} \frac{2}{\pi}\sum_{k=1}^{\infty} (-1)^{k+1} \frac{L}{l_{T}(\theta)}
\frac{\sin k(\phi_1-\phi_2)}{\sinh\frac{kL}{l_{T}(\theta)}}.
\label{current-density}
\end{eqnarray}
Here $l_T(\theta) = \frac{\hbar v_F \cos\theta}{2\pi k_BT} = \xi_T \cos\theta$.
 
In our assumptions (totally absorbing walls) the integration over
$\theta$ must be limited to the directions within the angle shown in
Fig.\ref{figY}, and the total Josephson current is written as  
\begin{eqnarray}
I(\phi_1-\phi_2)
 =  \frac{ev_F}{W\lambda_FL}\int\int_{-W/2}^{W/2} \frac{dy_1 dy_2}{\left[1+(\frac{y_1-y_2}{L})^2\right]^{3/2}}  \frac{2}{\pi}\sum_{k=1}^{\infty} (-1)^{k+1} \frac{L}{l_{T}(\theta_{y_1-y_2})}
\frac{\sin k(\phi_1-\phi_2)}{\sinh \frac{kL}{l_{T}(\theta_{y_1-y_2})}}
\end{eqnarray}
  (we take into account that $\tan\theta_{y_1-y_2} = (y_2-y_1)/L$).

The only difference that nonzero magnetic field will make is that
 $(\phi_1-\phi_2)$ in (\ref{current-density}) will be replaced by 
\begin{eqnarray} 
 \varphi({\bf n},{\bf r}) = \phi_2 - \phi_1 +
\frac{2\pi}{\Phi_0} \int^{0}_{\tau_1} 
{\bf A}({\bf r}-v_F\tau{\bf n})\cdot{\bf n}
d\tau,\end{eqnarray}
where the vector potential is integrated along the quasiclassical trajectory.
(We neglect here the dynamical effects of the magnetic field, small by the
parameter $\hbar\omega_c/\mu.$) The   expression for the  Josephson current becomes
 \begin{eqnarray}
I(\phi) =  \frac{ev_F}{W\lambda_FL}\int\int_{-W/2}^{W/2} \frac{dy_1 dy_2}{\left[1+(\frac{y_1-y_2}{L})^2\right]^{3/2}} 
 \frac{2}{\pi}\sum_{k=1}^{\infty} (-1)^{k+1} \frac{L}{\xi_{T}\cos\theta_{y_1-y_2}}
\frac{\sin k \left(\frac{\pi\Phi}{W\Phi_0}(y_1+y_2)+\phi\right)}{\sinh \frac{kL}{\xi_{T}\cos\theta_{y_1-y_2}}};
\label{tok}\\
I_c = \max_{0\leq \phi<2\pi} I(\phi).\label{tok2}
\end{eqnarray}
The current is  given by a sum of
the contributions from quasiclassical "Andreev tubes"  of width
$\sim\lambda_F$ each, following the quasiparticle trajectories.
Each contribution   depends on  the  length of the trajectory, and
 on the phase gained along  it (both from Andreev reflections at the NS
boundaries,
and from the vector potential on the way through the normal part of the system).
You will also notice that heuristic  formula (7) of Ref.\cite{Heida} 
correctly captures the qualitative picture of the effect and
  follows from   our expression in the limit
of narrow junction at high temperature ($L/W, L/\xi_T \to \infty$).

\begin{figure}
\epsfxsize=5in
\epsfbox{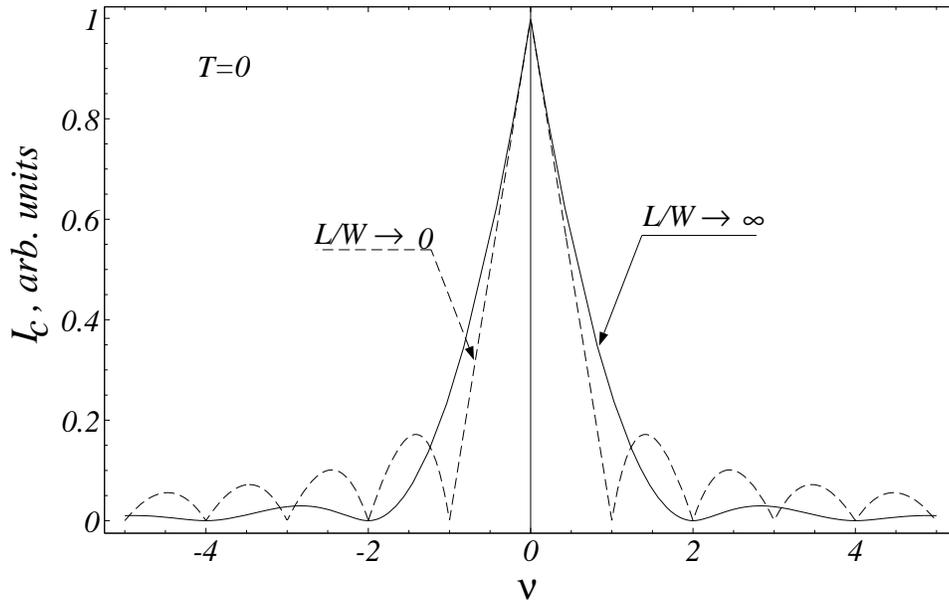}
\caption{Magnetic interference pattern at $T=0$ in wide ($L/W \to 0$) and narrow ($L/W \to \infty$)
 ballistic mesoscopic SNS junctions; $\nu = \Phi/\Phi_0.$}\label{figX}
\end{figure}

At zero temperature, the 
 results can  be obtained explicitly in the limiting cases of
wide ($L/W  \to 0$) and narrow ($L/W  \to \infty$) junctions.
Introducing $\nu =  \Phi/\Phi_0,$ we obtain (see Fig.\ref{figX})
\begin{eqnarray}
I_c(\nu) = \frac{2W}{\lambda_F}\frac{ev_F}{L} \frac{(1-\{\nu\})\{\nu\}}{|\nu|},\:\:L/W\to 0;\\
I_c(\nu) = \frac{W}{\lambda_F}\frac{ev_F }{ L } 
\frac{(1-\{\nu/2\})^2\{\nu/2\}^2}{|\nu/2|^2},\:\:L/W\to\infty,
\end{eqnarray}
where the first formula is  the standard result for a wide SNS junction;
all harmonics contribute to the current, producing the  distinctly 
 non-Fraunhofer picture \cite{Swidzinskii,SAB}. By $\{x\}$ we denote  the fractional part of $x$.

\begin{figure}
\epsfxsize=5in
\epsfbox{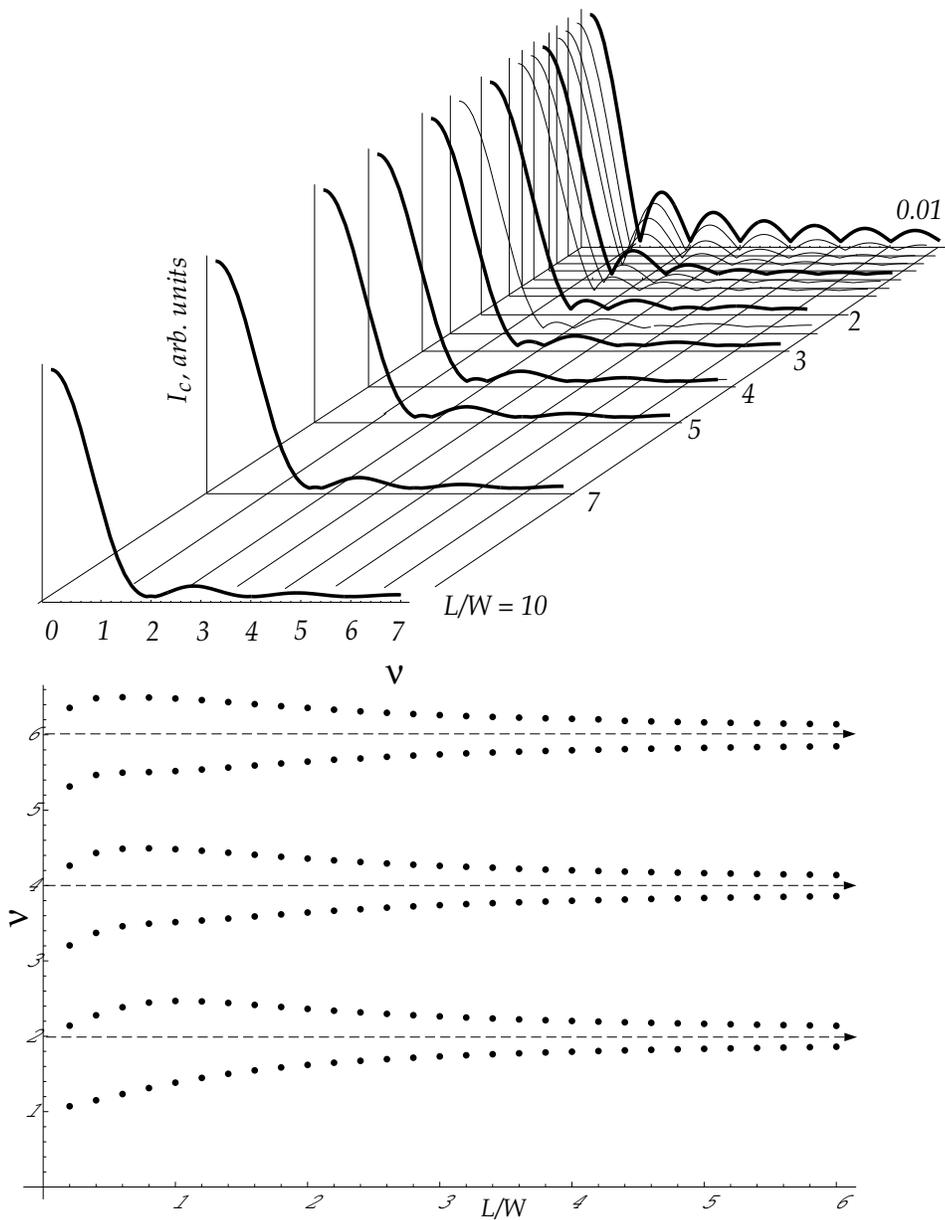}
\caption{Magnetic interference pattern and positions of zeros of the critical current as a
function of length-to-width ratio (at $L = 5\xi_T$). 
Note the nonmonotonic
behaviour of even zeros, and the fact that exact $\Phi_0$- or  $2\Phi_0$-
periodicity takes place only in the limiting cases of infinitely wide (narrow)
junction.}\label{figZ}
\end{figure}

\begin{figure}
\epsfxsize=5in
\epsfbox{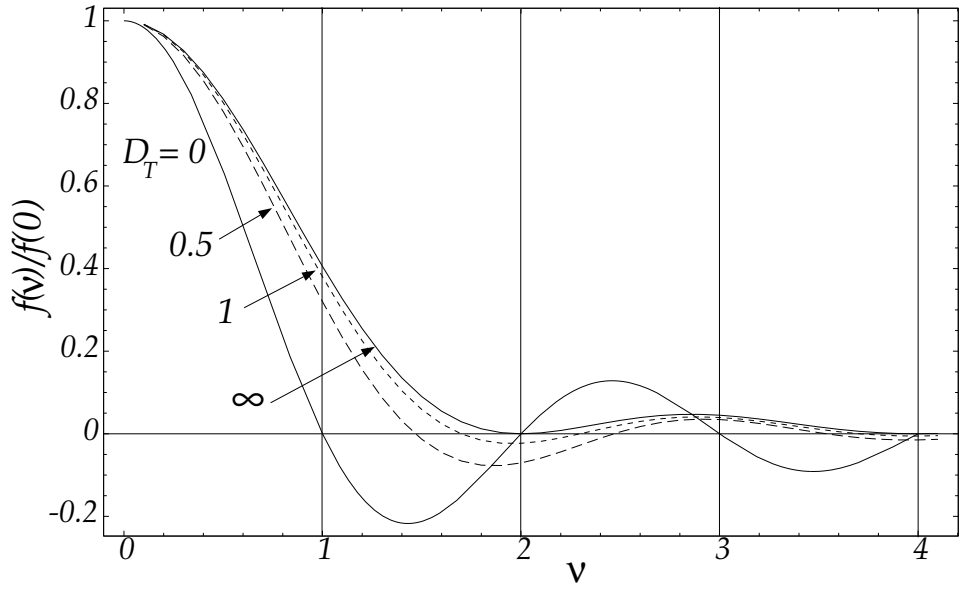}
\caption{Function $f(\nu)/f(0)$ at different values of $D_T = \sqrt{L\xi_T}/W$.
The critical current   $I_c(\nu) \propto |f(\nu)|.$
  Changing the temperature of the system of fixed size will lead to
evolution of the magnetic interference pattern.}\label{figH}
\end{figure}

In general case, the dependence $I_c(\nu)$ can be calculated numerically 
from (\ref{tok2}) (see 
Fig.\ref{figZ}). The period  is indeed smoothly evolving from 
$\Phi_0$ to $2\Phi_0$. Near the wide limit we see almost uniform growth
of the period, in agreement with our qualitative considerations, but in the 
intermediate regime, $L\sim W$, the  behaviour of   $I_c(\nu)$
becomes more complicated. It can be better understood in the high temperature 
limit, when only the   term with $k=1$ in (\ref{tok}) survives. Then the answer  can be obtained explicitly for an arbitrary value of  
$D = L/W$. Denoting by $D_T$ the "effective aperture",
\begin{eqnarray}
D_T = \frac{\sqrt{L\xi_T}}{W},\label{effective}
\end{eqnarray}
we can write the expression for the critical current as
\begin{eqnarray}
I_c(\nu) = \frac{2^{5/2} ev_F}{\pi^{3/2} \lambda_FD_T}
e^{-\frac{L}{\xi_T}} \left|f(\nu)\right|, \label{Ic} \\
f(\nu) = \nu^{-1} e^{-\frac{\pi^2D_T^2\nu^2}{2}}   {\rm Im} \left[e^{i\pi\nu}\left({\rm Erf}((1/D_T + i\pi D_T\nu)/\sqrt{2}) -
{\rm Erf}(i\pi D_T\nu/\sqrt{2})
\right) \right].\label{f}
%e^{-\frac{\pi^2D_T^2\nu^2}{2}}\times\nonumber\\
%\frac{\left|{\rm Im}\left[ 
%e^{i\pi\nu}\left({\rm Erf}((1/D_T + i\pi D_T\nu)/\sqrt{2}) -
%{\rm Erf}(i\pi D_T\nu/\sqrt{2})
%\right)\right]\right|}{|\nu|}.
\end{eqnarray}
The function $f(\nu)$
%\begin{eqnarray}
%\end{eqnarray}
 is proportional to $\sin(\pi\nu)/\nu$ at $L/W=0$, and becomes strictly 
positive (proportional to
  $\sin^2(\pi\nu/2)/\nu^2$) as  $L/W\to \infty$ (Fig.\ref{figH}). 
Thus the behaviour   of $I_c(\nu)$ near its zeros shown in Fig.\ref{figZ} is due to merging of real roots of $f(\nu)$.     
We see from (\ref{Ic},\ref{f}) that at finite temperature
 the  "wide-narrow" transition is governed by
$ 
\min(D_T,L/W).$
 Therefore another 
 way of observation of the effect is by changing the temperature of the sample
at fixed $L, W$. It has an advantage of being applicable to the mesoscopic
SNS junctions of fixed dimensions.

\begin{figure}
\epsfxsize=6in
\epsfbox{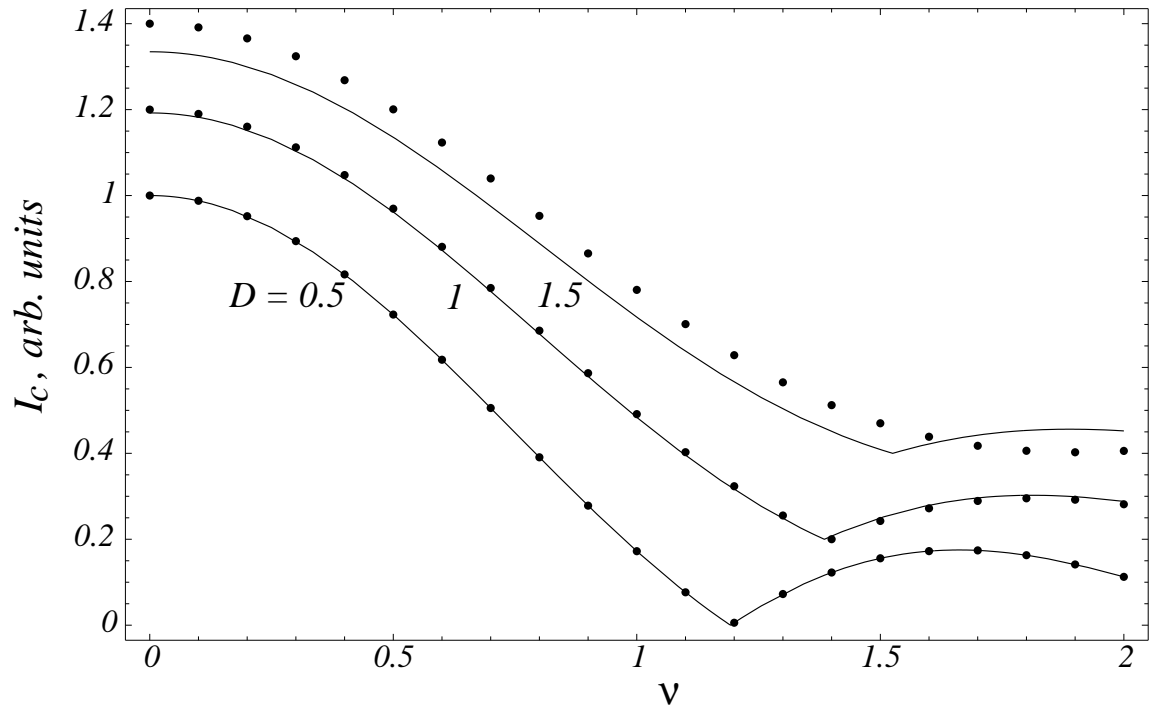}
\caption{Magnetic interference pattern at $L = 5\xi_T$ in case of total
absorption (solid line) and mirror reflection (dots) from
the side walls. The curves with different $D = L/W$
are offset in vertical direction.  }\label{figZZ}
\end{figure}

What difference would make a different boundary condition in the lateral
direction? In case of mirror reflection from the side walls the current is 
given by a simple generalization of (\ref{tok}), 
 including the contributions from the reflected 
trajectories. In the
absence of the magnetic field it is evident after we unfold
the reflected trajectories  by periodically extending  the system
in $y$-direction and taking the integral over $y_1$ from 
$-\infty$ to $\infty$, that the Josephson current 
per  width $W$ of the contact will be the same as in an infinitely wide
junction.
 But in the presence of the field the situation changes.  
Instead of linearly growing as $Hy$ (Eq.(\ref{gauge})), the effective vector
potential in the extended system will be periodic in $y$\cite{Aeff},
 and the field
contribution to the phase gain along the reflected trajectories will be
systematically less than along their counterparts in an infinitely wide 
junction. Therefore the qualitative picture of
   the evolution of the oscillation period remains valid,
as it is clearly seen in Fig.\ref{figZZ}.

\begin{figure}
\epsfxsize=7in
\epsfbox{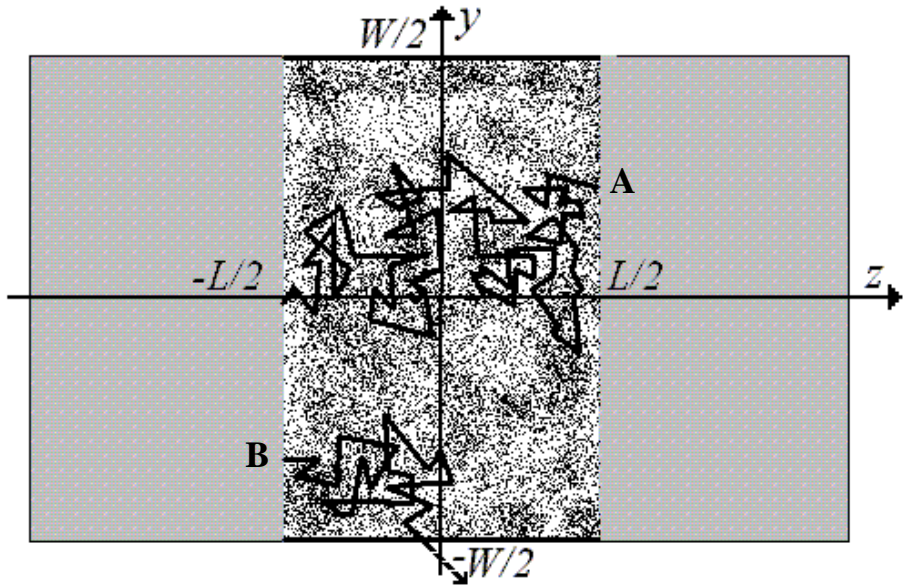}
\caption{Josephson current in an SNS junction in diffusive limit. Trajectory A does contribute to the Josephson current, trajectory B does not.}
\label{DIFFUSION.FIG}
\end{figure}

   Effect of strong elastic scattering in the normal layer is more drastic. Let us consider the limiting case of dirty SNS junction (Fig.\ref{DIFFUSION.FIG}). Now the system is  described by the Usadel equations for the quasiclassical Green's functions  averaged over directions\cite{Usadel}, $F_{\omega}$ and $G_{\omega}$. These equations and the expression for the supercurrent density read \cite{Swidzinskii}(from now on we put $\hbar = c = 1$): \begin{eqnarray} |\omega| F_{\omega}({\bf r}) + \frac{D}{2}\left( F_{\omega}({\bf r})\nabla^2 G_{\omega}({\bf r}) - G_{\omega}({\bf r}) \left(\nabla+2ie{\bf A}({\bf r})\right)^2 F_{\omega}({\bf r}) \right) = \Delta^*({\bf r});\\|F_{\omega}|^2 + G_{\omega}^2 = 1;\\  {\bf j}({\bf r}) = -\pi e {\cal  N}(0) D T \sum_{\omega} {\tt Im} \left(F_{\omega}^*({\bf r}) (\nabla+2ie{\bf A}({\bf r}))F_{\omega}({\bf r})\right). \label{TOK-Uzi} \end{eqnarray}
  
We will limit our considerations to long enough junctions and high enough temperatures, $L > \tilde{\xi}_T.$ Then the sum over Matsubara frequencies in (\ref{TOK-Uzi}) will be dominated by the terms with $|\omega| = \pi k_B T$;  the anomalous Usadel function will be small inside the normal layer. Therefore the first Usadel equation  can be  linearized to yield \begin{equation} q_T^2F_{\omega}({\bf r}) - \left(\nabla+2ie{\bf A}({\bf r})\right)^2 F_{\omega}({\bf r})  = 0, \label{LINEAR-Uzi} \end{equation} where $q_T^2 = 2|\omega|/D = 1/\tilde{\xi}_T^2$.  The boundary conditions for $F_{\omega}({\bf r}) $ at $z=\pm L/2$ and $y=\pm W$  can be chosen as\cite{Swidzinskii} \begin{equation} F_{\omega}(y,-L/2) = f_0(\omega) ; \:\:\: F_{\omega}(y,L/2) = f_0(\omega) e^{i(\phi_2-\phi_1)}, \end{equation} where $f_0(\omega)  = \Delta_{\pm L/2}/(|\omega|+\sqrt{|\omega|^2+ \Delta_{\pm L/2}}).$ We are not interested in the actual value of the order parameter on the NS boundary (which due to proximity effect is lower than in the bulk), as long as the anomalous Green's function stays small. We choose zero boundary conditions at $y=\pm W/2$, which would correspond to e.g. contact with clean normal conductor; the Josephson current is again carried only by the quasiparticles whose trajectories link    the superconductors (Fig.\ref{DIFFUSION.FIG}).
After eliminating the vector potential from (\ref{TOK-Uzi},\ref{LINEAR-Uzi}) by the transformation $F_{\omega}({\bf r})= \tilde{F}_{\omega}({\bf r}) e^{-2ie{\bf A}{\bf r}}$,   boundary conditions at $\pm L/2$ are modified:\begin{equation} \tilde{F}_{\omega}(y,-L/2) = f_0(\omega)e^{-i\pi\nu y} ; \:\:\: \tilde{F}_{\omega}(y,L/2) = f_0(\omega) e^{i(\phi_2-\phi_1)}e^{i\pi\nu y} .\end{equation} The solution is readily expressed through the Green's function of Eq.(\ref{LINEAR-Uzi}), ${\cal G}(y,z;\eta,\zeta)$,     in the rectangle with zero boundary conditions\cite{Madelung}: \begin{eqnarray*} \tilde{F}_{\omega}(\eta,\zeta) =   \frac{f_0(\omega)}{2\pi}\int_{-W/2}^{W/2} dy \left( e^{-i\pi\nu y} \left.\frac{\partial}{\partial z}{\cal G}(y,z;\eta,\zeta)\right|_{z=-L/2} - e^{i(\phi_2-\phi_1)}e^{i\pi\nu y}  \left.\frac{\partial}{\partial z} {\cal G}(y,z;\eta,\zeta)\right|_{z=L/2} \right). \end{eqnarray*}  The latter can be found by the method of mirror images:  \begin{equation} {\cal G}(y,z;\eta,\zeta) = \sum_{m=-\infty}^{\infty} \sum_{n=-\infty}^{\infty}  (-1)^m (-1)^n {\cal G}^0(mW+(-1)^m y,nL+(-1)^n z;\eta,\zeta). \end{equation} Here ${\cal G}^0(y,z;\eta,\zeta) = \frac{1}{2\pi}K_0(q_T\sqrt{(y-\eta)^2+(z-\zeta)^2})$ is  the Green's function in the infinite plane,   $K_0$ is the modified Bessel function. 

\begin{figure}
\epsfxsize=5in
\epsfbox{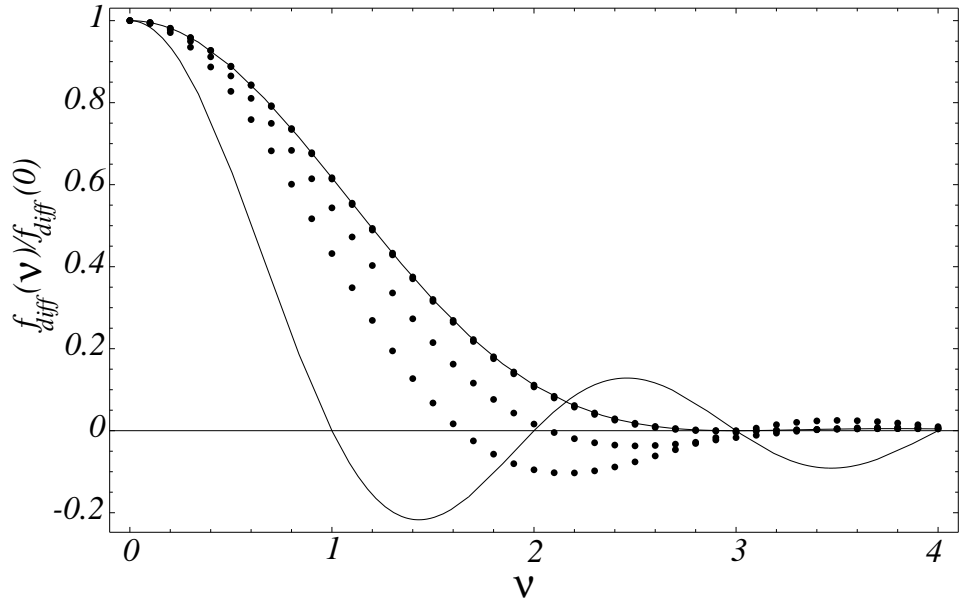}
\caption{Function $f_{diff}(\nu)/f_{diff}(0)$ at $L=2\tilde{\xi}_T$
and different values of $W/\tilde{\xi}_T = 0.5; 1; 2; 3$ (dots). Solid lines  show the limiting cases, $\cos^2(\pi\nu/2)/(1-\nu^2)^2$ and $\sin (\pi\nu)/(\pi\nu)$ respectively. 
The critical current in a dirty SNS junction  $I_c(\nu) \propto |f_{diff}(\nu)|.$}
\label{FIG.TRIPLE}
\end{figure}

Summation over $m$ can be limited to $m=0,\pm 1$ because of our assumption $L > \tilde{\xi}_T$. The sum over $n$ is taken using the Poisson summation formula.   Substituting this in the expression for the current density at $z=0$, we finally find for the Josephson
current: $$ I(\phi_2-\phi_1,\nu) = I_c(\nu) \sin(\phi_2-\phi_1),$$ where \begin{eqnarray} I_c(\nu) = \pi e {\cal N}(0) D T |f_0|^2 \left|f_{diff}(\nu)\right|,\\ f_{diff}(\nu) = \sum_{l=-\infty}^{\infty} (-1)^l S_l(L/2)S_l'(L/2) \left(\frac{\sin \pi(\nu+l)/2}{\pi(\nu+l)/2} - (-1)^l\frac{\sin \pi(\nu-l)/2}{\pi(\nu-l)/2}\right)^2,\\                  S_l(u) =   \sqrt{|u|/2\pi} (q_T^2+\pi^2l^2/W^2)^{1/4}K_{1/2}(\sqrt{u^2(q_T^2+\pi^2l^2/W^2)});\:\:\:                    
S_l'(u) = \frac{d}{d u}S_l(u).  \end{eqnarray} Behaviour of the function $f_{diff}(\nu) $ is shown in Fig.\ref{FIG.TRIPLE}. Due to exponential decay of $S_l(L/2)$ and its derivative, in the limit of narrow junction ($W \ll \tilde{\xi}_T,L$) only the terms with smallest $l = \pm 1$ survive\cite{MATS}. This yields $f_{diff}(\nu)/f_{diff}(0) = \cos^2(\pi\nu/2)/(1-\nu^2)^2$. The first zero of this function is at $\Phi = 3 \Phi_0$, {\em tripling} the first  oscillation period. The following zeros are separated by $2\Phi_0$-intervals
(though the amplitude of higher-order peaks decreases so fast with $\nu$, that the possibility to observe this periodicity is rather problematic). In the opposite limit of wide junctions, $W\gg\tilde{\xi}_T,$  $S_l$'s cease to depend on $l$, and the shape of the interference pattern will be given by $$\sum_{l=-\infty}^{\infty} (-1)^l \left(\frac{\sin \pi(\nu+l)/2}{\pi(\nu+l)/2} - (-1)^l\frac{\sin \pi(\nu-l)/2}{\pi(\nu-l)/2}\right)^2 \equiv -4\frac{\sin\pi\nu}{\pi\nu},$$  the Fraunhofer pattern expected in a wide dirty SNS junction.
As in the ballistic case, continious change of the periodicity can be caused  by lowering the temperature of the system.  

Note that in the range of parameters $(W \sim L \sim \tilde{\xi}_T)$ when the first zero of the critical current
appears at $\nu \sim 2 $ the curve $I_c(\nu) ( \propto |f_{diff}(\nu)|)$ is 
distinctly sharper than in the ballistic case (see Figs.\ref{figH},\ref{FIG.TRIPLE}). Since
 the measurements of Ref.\cite{Heida} (with diffusive scattering at NS boundaries) produced sharp cusps rather than smooth minima at $\Phi \sim 2\Phi_0$, one can speculate that their system was   effectively in
diffusive rather than ballistic regime.

Qualitatively the effect can also be understood in terms of loss of available phase space for current-carrying quasiparticles.  Only the diffusive trajectories linking both superconductors contribute to the Josephson current (carried along such a trajectory by electrons   and holes related through Andreev reflections at the end points).  Therefore the trajectories originating too close to the lateral boundaries have more chances to end on the boundary and be lost, and the effective width of the junction, as well as the effective magnetic flux through it, are less than their actual values. Finite width of the junction also limits the "wandering'' trajectories with large effective phase gain. Due to strong elastic scattering, the average $z$-component of the current carried through such a trajectory is the same as for a "shortcut" one, quite unlike the ballistic case where contributions from "grazing'' trajectories are systematically less. This  explains why the periodicity change is even more drastic in the diffusive case.

In conclusion, we have demonstrated that the periodicity
 of the critical current dependence on the 
magnetic flux penetrating a mesoscopic SNS junction generally  
depends on the geometry of the system and is not given by either of
fundamental quanta, $\Phi_0 = hc/2e$ or   $2\Phi_0 = hc/e$, 
in a  contrast to
the case of tunneling Josephson junction. 
 The effect
is a manifestation of the nonlocal   electrodynamics in hybrid normal-superconducting
systems on
scale $\xi_T (\tilde{\xi}_T) \gg \xi_0$. It  can be understood in  terms of quasiclassical 
Andreev levels  ("Andreev tubes"), following the  quasiparticle
trajectories in the normal part of the system. Our theory provides
an explanation for the recent experimental observation of $2\Phi_0$-periodic 
magnetic interference pattern in S-2DEG-S junctions \cite{Heida},
 and predicts a continuous
transition from $\Phi_0$ to $2\Phi_0$ periodicity in ballistic case, and to $2\Phi_0$-periodicity with the first iscillation period $3\Phi_0$ on the dirty limit.  This transition can be 
observed either by changing the width of the 2DEG layer in the split-gate
technique, or by changing the temperature of the fixed-size junction. The latter
approach can be applied as well to mesoscopic SNS junctions  with normal metal layer instead of 2DEG.

We wish to thank I. Affleck, J.-S. Caux, L.P. Gor'kov, U. Ledermann, and P.C.E. Stamp for valuable discussions. This work was supported by NSERC of
Canada and in part by CIAR (AZ) and by the NHMFL
through NSF cooperative agreement No. DMR-9527035 and the
State of Florida (VB).

\newpage

\newpage

\newpage

\newpage

\newpage

\newpage

\newpage

  \end{document}